\documentclass[prd,twocolumn,superscriptaddress,showpacs,preprintnumbers,amsmath,amssymb,APS]{revtex4-1}
\usepackage{bm}
\usepackage[latin1]{inputenc}
\usepackage[spanish,english]{babel}
\usepackage{amsfonts}
\usepackage{amssymb}
\usepackage{graphicx}
\usepackage{color}

\newcommand{\be}{\begin{equation}}
\newcommand{\ee}{\end{equation}}
\newcommand{\bea}{\begin{eqnarray}}
\newcommand{\eea}{\end{eqnarray}}

\begin{document}

\title{{\bf Electric-magnetic duality and renormalization in curved  spacetimes}}

\author{Ivan Agullo}\email{agullo@lsu.edu}\affiliation{Department of Physics and Astronomy, Louisiana State University, Baton Rouge, LA 70803-4001;}
\author{Aitor Landete}\email{aitor.landete@csic.es}
\affiliation{Instituto de Fisica Teorica UAM-CSIC, Universidad Autonoma de Madrid, Cantoblanco, 28049 Madrid, Spain.}
\affiliation{Department of Physics and Astronomy, Louisiana State University, Baton Rouge, LA 70803-4001;}
\affiliation{Departamento de Fisica Teorica and IFIC, Centro Mixto Universidad de Valencia-CSIC. Facultad de Fisica, Universidad de Valencia, Burjassot-46100, Valencia, Spain.}\author{Jose Navarro-Salas}\email{jnavarro@ific.uv.es}
\affiliation{Departamento de Fisica Teorica and IFIC, Centro Mixto Universidad de Valencia-CSIC. Facultad de Fisica, Universidad de Valencia, Burjassot-46100, Valencia, Spain.}

\date{November 27, 2014}

\begin{abstract}

We point out that the  duality symmetry of  free electromagnetism does not hold  in the quantum theory if an arbitrary classical gravitational background is present. The symmetry breaks in the process of renormalization, as also happens with conformal invariance. We  show that a similar duality anomaly appears   for a  massless scalar field in $1+1$ dimensions.

\end{abstract}


\maketitle


\section {Introduction and summary} 
The  Maxwell equations {\em in vacuo} are highly symmetric. In addition to  their relativistic (Poincaré) invariance in Minkowski spacetime, they exhibit two additional symmetries: conformal---or Weyl---invariance and electric-magnetic duality. 
The former is the symmetry under  Weyl  transformations (or conformal re-scalings) $g_{\mu\nu}\to \Omega^2(x)g_{\mu\nu}$ \cite{note1}. This is a symmetry of the classical theory in arbitrary spacetimes, and it is also an exact symmetry of the quantum theory in Minkowski spacetime. However, as first pointed out in \cite{capper-duff}, Weyl invariance cannot be extended to quantum field theory (QFT) in {\it curved} backgrounds. Weyl symmetry implies the tracelessness of the energy-momentum tensor $T_{\mu \nu}$. Since $T_{\mu \nu}$ is quadratic in the field $F_{\mu\nu}$, renormalization is required to compute its expectation values. It turns out that  generally covariant methods of renormalization in curved spacetime produce a nonvanishing trace  $\langle T^{\mu}_{\mu}\rangle$ which breaks the Weyl invariance  \cite{birrell-davies,Waldbook,parker-toms}. The value of this trace is independent of the state in which the expectation value is evaluated, and is written in terms of curvature tensors. The breakdown of Weyl symmetry is a renormalization effect and therefore it  is only manifest when composite operators are considered, such as  $T_{\mu\nu}$ (the equations of motion and correlation functions are still Weyl invariant). This is the celebrated Weyl or  trace anomaly (also called the conformal anomaly), which constitutes a robust prediction of renormalization in curved spacetimes and has important physical consequences \cite{birrell-davies,Waldbook,parker-toms}. 

Another important symmetry of electromagnetism in the absence of charges is invariance under duality transformations $F_{\mu\nu} \to {^*F}_{\mu\nu}$ (see e.g. \cite{jackson, deser-teitelboim}) where the (Hodge) dual tensor is defined in the standard way,  ${^*F}^{\mu\nu}= 1/2|g|^{-1/2}\epsilon^{\mu\nu\alpha\beta}F_{\alpha\beta}$. In terms of the electric and magnetic fields, this discrete transformation reads $\vec E \to \vec B, \vec B \to - \vec E$. It can be also viewed as a particular case of the electric-magnetic rotation  $\vec{E} \to \vec{E} \cos{\theta} +\vec{B} \sin{\theta}$,  $\vec{B} \to \vec{B} \cos{\theta} -\vec{E} \sin{\theta}$. Maxwell's equations $$\nabla_{\mu}F^{\mu\nu}=0\, , \, \, \nabla_{\mu}  {^{*}F}^{\mu\nu}=0\, ,$$ are manifestly invariant. The classical (Maxwell) energy-momentum tensor, which can be written in the symmetric form
\be \label{TM}T_{\mu\nu}^{\rm M} = -\frac{1}{2}( F_{\mu\alpha}F_{\nu}^{\ \alpha}+ {^{*}F}_{\mu\alpha} {^{*}F}_{\nu}^{\ \alpha}) \ ,  \ee
is also invariant. This duality  can be extended to the QFT in Minkowski spacetime. One can show that the duality transformation is implemented by a unitary operator in the Fock space which leaves  the Minkowski vacuum invariant. As a consequence, vacuum-correlation functions are duality invariant, e.g.\ $$\langle F_{\mu\alpha}(x)F_{\nu}^{\ \alpha}(x') \rangle = \langle  {^{*}F}_{\mu\alpha}(x) {^{*}F}_{\nu}^{\ \alpha}(x')\rangle$$ for all $x\neq x'$. 
Vacuum  expectation values of composite (nonlinear) operators are also invariant, although renormalization is required to make sense of the otherwise divergent expressions. In Minkowski spacetime normal order (i.e.\ subtraction of the vacuum expectation value)  does the job. As an example, one trivially  obtains  $\langle  F_{\mu\alpha}(x)F_{\nu}^{\ \alpha}(x) \rangle = \langle  {^{*}F}_{\mu\alpha}(x) {^{*}F}_{\nu}^{\ \alpha}(x)\rangle=0$.

The goal of this paper is to show that the classical electric-magnetic duality symmetry cannot be extended to QFT in {\em curved} spacetime. In order to show the influence of the  gravitational background in the sharpest way, we will work as closely as possible to the theory in Minkowski spacetime. We will consider free electromagnetism, $\mathcal{L}=-1/4 \sqrt{|g|} F^{\mu\nu}F_{\mu\nu}$,  in a spatially flat Friedman-Lemaitre-Robertson-Walker (FLRW) spacetime. This background is {\em conformally Minkowskian} and, since the electromagnetic field equations are Weyl invariant, the quantum theory shares multiple properties with the Minkowski spacetime formulation. In particular, both theories have the same Hilbert space. This relation allows the definition of a {\em preferred vacuum state} in FLRW backgrounds (the so-called conformal vacuum), and also implies the absence of particle (i.e.\ photon) creation by the expanding spacetime, in sharp contrast with other non-Weyl invariant fields \cite{parker-toms}. However, the presence of a nontrivial spacetime curvature manifests itself in an important way in the  process of renormalization. Although there exists a preferred vacuum, the normal order prescription  is not a satisfactory renormalization prescription in FLRW. This is because that procedure for subtracting the ultraviolet divergences  is neither generally covariant nor local. Therefore,  out of Minkowski spacetime, normal-order does not satisfy the axioms on which the theory of renormalization in curved spacetime relies \cite{Waldbook}.  Instead, we will use the adiabatic renormalization  method developed by Parker and Fulling \cite{parker-fulling, parker-toms, birrell-davies}---which has been proven to be equivalent to  DeWitt-Schwinger point-splitting renormalization \cite{Christensen, birrell-davies}---adapted to the electromagnetic field. (See also \cite{chimento-cossarini}, and see  \cite{landete-navarro-torrenti} for the extension to fermionic fields.) We will show that the quantity 
\be \label{deltadef} \Delta_{\mu\nu} \equiv \langle F_{\mu\alpha}(x)F_{\nu}^{\ \alpha}(x) \rangle - \langle  {^{*}F}_{\mu\alpha}(x) {^{*}F}_{\nu}^{\ \alpha}(x)\rangle \ee
takes a  nonvanishing value  given (we use the same geometric conventions as in Refs. \cite{parker-toms, birrell-davies}) by
\be \label{Delta}\Delta_{\mu\nu}= \frac{1}{480  \pi^2} \left(-\frac{9}{2}R_{\alpha \beta}R^{\alpha \beta} + \frac{23}{12}R^2 +2  \Box R\right) \, g_{\mu\nu} \, ,\ee
where $R_{\alpha \beta}$ is the Ricci tensor and $R$ its trace. This expression implies that the  fluctuations of the electric and magnetic field  in the  vacuum state---which {\em is}  duality invariant in FLRW---are different, i.e.\  $\langle \vec E^2 \rangle \neq \langle \vec B^2 \rangle$, and therefore the duality symmetry is broken. 

We analyze  the same issue in the case of a massless, minimally coupled scalar field in an arbitrary $1+1$-dimensional spacetime. We use the Hadamard renormalization  method and reach  similar conclusions: the presence of a classical gravitational background breaks not only Weyl  invariance, but also the duality symmetry. \\

\section{Duality anomaly for the electromagnetic field}

 The goal of this section is to compute the vacuum expectation values $\langle F_{\mu\alpha}(x)F_{\nu}^{\ \alpha}(x) \rangle$,  $\langle  {^{*}F}_{\mu\alpha}(x) {^{*}F}_{\nu}^{\ \alpha}(x)\rangle$, and  the energy-momentum tensor  in a spatially flat FLRW background with line element $$ds^2= a(\eta)^2 (d\eta^2-d\vec{x}^2)\, ,$$ where $\eta$ is the conformal time. All tensor components in this section will refer to the coordinates $\eta,\vec{x}$. As pointed out above, the conformal invariance of the equations of motion greatly facilitates the formulation of the theory. The electromagnetic field operator can be written  in terms of the vector potential as $ F_{\mu\nu}=\nabla_\mu A_{\nu}-\nabla_\nu A_{\mu}$, where the operator $A_{\mu}$ can be represented in terms of Fourier modes of the two physical polarizations (we work in the Lorenz gauge $\nabla^{\mu}A_{\mu}=0$) by
\be \label{A} A_{\mu}(\vec{x},\eta)=\int \frac{d^3k }{(2\pi)^3} \, \sum_{\alpha=1}^2 \, \hat a^{(\alpha)}_{\vec{k}} \, \epsilon^{(\alpha)}_{\mu} \,  \varphi_{\vec{k}}(\eta)  \, e^{i \vec{k}\vec{x}}+\, {\rm H.c.} \ee
where $\varphi_{\vec{k}}(\eta)=e^{-i k\eta}/\sqrt{2 k}$, $k=|\vec{k}|$ is the length of the comoving mode $\vec{k}$, and   $\hat a^{(\alpha)}_{\vec{k}}$ and $\hat a^{(\alpha)\, \dagger}_{\vec{k}}$ are creation and annihilation operators for the polarization $\alpha$. The polarization vectors $\epsilon^{(\alpha)}_{\mu}(\vec{k})$  depend on $\vec{k}$ and are transversal, $k^{\mu}\epsilon^{(\alpha)}_{\mu}=0$. It is convenient to choose them to be mutually orthogonal, $g^{\mu\nu} \epsilon^{(\alpha)}_{\mu} \epsilon^{(\alpha')}_{\nu}=-a^{-2} \, \delta^{\alpha \alpha'}$. 

Direct substitution shows that the quantity $\langle F_{\mu \sigma}{F_{\nu}}^{\sigma}\rangle$ is ultraviolet divergent, and therefore requires renormalization. In adiabatic renormalization the physically relevant, finite expression is obtained by subtracting mode by mode, i.e.\ under the Fourier integral sign, terms that would  produce state-independent ultraviolet divergences. The terms to be subtracted are identified by performing a Liouville or WKB-type asymptotic expansion for large values of the physical frequency of the Fourier modes $\omega(k)$ or, mathematically equivalently, an expansion for small values of the time derivatives of the scale factor $a(\eta)$ (this is the reason for the name adiabatic, although the method is primarily concerned with ultraviolet issues). See \cite{parker-toms} for further details. 

A lengthy calculation produces (see Appendix A for details)
\be\langle  {F}_{\mu\alpha}{F}_{\nu}^{\ \alpha}\rangle= \theta_{\mu\nu} +\frac{1}{4}\gamma(\eta) \, g_{\mu\nu} +{^{}t}_{\mu\nu} \label{FF} \, , \ee
\be\langle  {^{*}F}_{\mu\alpha}{^{*}F}_{\nu}^{\ \alpha}\rangle= \theta_{\mu\nu} -\frac{1}{4}\gamma(\eta) \, g_{\mu\nu}  +{^{*}t}_{\mu\nu} \, .\label{*F*F} \ee
In these expressions  ${^{}t}_{\mu\nu}$ and ${^{}*t}_{\mu\nu}$ are traceless tensors encoding all the information regarding the quantum state, and both vanish for the conformal vacuum. $\theta_{\mu\nu}$ is a traceless, local geometric tensor given by 
\bea   \theta_{\mu\nu}&=&\frac{1}{480 \pi^2}\Big[\frac{-16}{3}R_{\mu \alpha}R^{\alpha}_{\nu} + \frac{61}{18}RR_{\mu\nu} +\frac{2}{3}\nabla_{\nu} \nabla_{\mu}R\nonumber \\ &+&\frac{4}{3}R_{\alpha\beta}R^{\alpha\beta}g_{\mu\nu}-\frac{61}{72}R^2 g_{\mu\nu}  -\frac{1}{6}\Box R g_{\mu\nu}\Big]  \, \eea 
and $$\gamma(\eta)= \frac{1}{ 480 \pi^2}[-9 R_{\alpha\beta}R^{\alpha\beta}+\frac{23}{6}R^2  +4\Box R]\, .$$ Applying Eqs.\ (\ref{FF}) and (\ref{*F*F}) for the vacuum state,  one obtains  $\Delta_{\mu\nu}=1/2\, \gamma(\eta)\, g_{\mu\nu}$, as anticipated in Eqs.\ (\ref{deltadef}) and (\ref{Delta}). This quantity is different from zero for a generic scale factor $a(\eta)$. Also note that taking the trace of Eq.\ (\ref{FF}) one obtains $\langle {F}^2\rangle\equiv \langle {F}_{\mu\alpha}{F}^{\mu\alpha} \rangle= \langle \vec E^2 \rangle - \langle \vec B^2 \rangle =\gamma(\eta)$. Since the vacuum state is duality invariant, these results indicate a breakdown of duality. 

As in Minkowski spacetime, in FLRW there also exists a unitary operator implementing the duality transformation in the  representation of the  (linear) Heisenberg algebra of field operators. However, the previous result indicates that the renormalized expectation values of composite (nonlinear) operators do not transform as expected under this unitary operator. The geometric quantities---curvature tensors---involved in the renormalization procedure break the duality symmetry. 

We finish this section by providing the expression for the renormalized energy-momentum tensor. The aim  is to show that our techniques are consistent with well-known results  on  curved-space renormalization of the electromagnetic field, which maintain general covariance  and gauge invariance \cite{Christensen, Adler77}.  From expression  (\ref{TM}) the vacuum expectation value of the Maxwell tensor is
\be  \langle T_{\mu\nu}^{\rm M}(x) \rangle= -\frac{1}{2}( \langle  {F}_{\mu\alpha}{F}_{\nu}^{\ \alpha}\rangle + \langle  {^{*}F}_{\mu\alpha}{^{*}F}_{\nu}^{\ \alpha}\rangle)=-\theta_{\mu\nu}\, .\ee
 By construction, this tensor is traceless. However, $\langle T_{\mu\nu}^{\rm M}(x) \rangle$ is not a suitable candidate for the source of the gravitational field, i.e.\  for the right-hand side of the semiclassical  Einstein equations $G_{\mu\nu}=- 8\pi G\, \langle T_{\mu\nu}\rangle$, since $\langle T_{\mu\nu}^{\rm M}(x) \rangle$  is {\em not  conserved}, $\nabla^{\mu}\langle T_{\mu\nu}^{\rm M}(x) \rangle\neq 0$. Explicit computations show that
\be \label{nablaTM} \nabla^{\mu}\langle  T^{\rm M}_{\mu\nu} \rangle= -\frac{1}{2 (4\pi)^2} [\nabla^{\mu} v_{\mu\nu}-\frac{3}{4} \nabla_{\nu}v^{\rho}_{\rho}+\nabla_{\nu} v]\ee
where $v_{\mu\nu}$ and $v$ are objects constructed from  curvature tensors:
\bea  v_{\mu\nu}&=&\frac{1}{3}R_{\mu \alpha}{R^{\alpha}}_{\nu} - \frac{3}{10}RR_{\mu\nu} 
- \frac{1}{45}\nabla_{\nu} \nabla_{\mu}R \\\nonumber   &+&\frac{1}{180}R_{\alpha\beta}R^{\alpha\beta}g_{\mu\nu}+\frac{113}{2160}R^2 g_{\mu\nu}-\frac{1}{360}\Box R g_{\mu\nu} \, , \eea
\be v=\frac{13}{1080}R^2+\frac{1}{30} \Box R+\frac{1}{180} R^{\alpha\beta}R_{\alpha\beta}\, .\ee
 One can construct a suitable conserved energy-momentum tensor from $\langle T_{\mu\nu}^{\rm M}(x) \rangle$ in two different ways. The shortest one is to use the procedure commonly employed in Hadamard renormalization \cite{Adler77, wald78, Waldbook, Hadamard}. It consists  of simply adding to $\langle  T^{\rm M}_{\mu\nu} \rangle$ a geometric tensor that makes it conserved. From Eq.\ (\ref{nablaTM}) we find that a solution is
\be \label{Tmn} \langle  T_{\mu\nu} \rangle =   \langle  T^{\rm M}_{\mu\nu} \rangle +  \mathcal{T}^{\rm Ad}_{\mu\nu} + c_1 H^{(1)}_{\mu\nu}\ . \ee
 where  $$\mathcal{T}^{\rm Ad}_{\mu\nu}=\frac{1}{2 (4\pi)^2} [v_{\mu\nu}+(-\frac{3}{4} v^{\rho}_{\rho}+ v) g_{\mu\nu}]\, .$$ Of course, this method can only define $ \langle  T_{\mu\nu} \rangle $ up to a conserved tensor. In FLRW this ambiguity is all encoded in the last term of the previous equation, where $c_1$ is an arbitrary real number and  $H^{(1)}_{\mu\nu}$ is the tensor  obtained by functional variation of $\sqrt{-g}R^2$ with respect to the metric; therefore it is conserved, $\nabla^{\mu}H^{(1)}_{\mu\nu}=0$. Note that the freedom in the value of $c_1$ in (\ref{Tmn}) coincides with the well-understood ambiguity in the renormalized energy-momentum tensor in curved spacetimes \cite{wald78,Waldbook}.

Another way of finding $\langle  T_{\mu\nu} \rangle $ in adiabatic renormalization is by direct application of the method. But to follow this route one has to deal carefully with the gauge invariance. A convenient approach in curved backgrounds is provided by  the Faddeev-Popov scheme (see e.g.\ \cite{birrell-davies}). This method introduces new contributions to the energy-momentum tensor, namely the so-called  gauge breaking terms and the contribution of a ghost field. Explicit computations produce results that agree with  (\ref{Tmn}).

From (\ref{Tmn}) it is easy to check that the trace of the renormalized energy-momentum tensor is nonzero and is given by  
\bea \langle  T^{\mu}_{\mu} \rangle&=& \frac{1}{2880 \pi^2}[-62(R^{\alpha\beta}R_{\alpha\beta}-\frac{1}{3} R^2)\nonumber\\&-&(2+ 6 \times 2880\pi^2\, c_1) \, \Box R]\, .\eea This is the well-known  trace anomaly.  Any other renormalization method would provide an expression for $ \langle  T_{\mu\nu} \rangle$ that would possibly differ from (\ref{Tmn}) in the value of the coefficient $c_1$.
Note that the existence of the anomalous trace  does not imply the duality anomaly. The trace  arises from the geometric term $\mathcal{T}^{\rm Ad}_{\mu\nu}$ in (\ref{Tmn}), while the duality anomaly appears already in the expectation values (\ref{FF}) and (\ref{*F*F}).\\

\section{Results in de Sitter universe}
For the de Sitter-FLRW solution  [$a(t)= e^{-Ht}$ in cosmic time and $a(\eta)=-1/(H\eta)$ in conformal time, with $dt= a \, d\eta$] the conformal vacuum---also called the Bunch-Davies vacuum---is de Sitter invariant. Evaluation of Eq.\ (\ref{TM}) produces $\langle T^M_{\mu\nu} \rangle =0$.  This result is expected from symmetry arguments, since there are no two-covariant  tensors which are simultaneously de Sitter invariant   and traceless.
If one (incorrectly) assumes, following the standard lore, the validity of electric-magnetic duality (see e.g.\ \cite{Adler77,agullo-navarro}), one would conclude that $\langle \vec B^2 \rangle =0= \langle \vec E^2 \rangle$ in this spacetime \cite{agullo-navarro}.  However, particularizing Eqs.\ (\ref{*F*F}) and  (\ref{FF}) to  de Sitter space one obtains, instead, $\langle \vec B^2 \rangle = \frac{19}{160 \pi^2} H^4$---in  agreement with \cite{campanelli}---and $\langle \vec E^2 \rangle = -\langle \vec B^2 \rangle$. The negative value of the quadratic quantity $\langle \vec E^2 \rangle$ should not be surprising since that is common for renormalized quantities. The same happens in the usual Casimir effect (see e.g.\ \cite{DeWitt75}).\\

\section{Duality anomaly in a $2D$ conformal scalar theory\label{2D}}
The duality anomaly in curved spacetimes can also be illustrated in a simpler scenario: a minimally coupled, massless scalar field in $1+1$ dimensions. This theory is very similar to free electromagnetism in the sense that it can be described by an Abelian $1$-form $F_{\mu}$ \cite{Deser-Gomberoff}, and classically it shows both Weyl and duality invariance. This framework has been extensively discussed in the context of conformal field theory and string theory \cite{note2}.

The classical stress-energy tensor can be expressed as 
$$T_{\mu\nu} = \frac{1}{2}(F_{\mu}F_{\nu} + {^{*}F}_{\mu} {^{*}F}_{\nu})\, ,$$ 
where ${^{*}F}_{\mu} = |g|^{1/2}\epsilon_{\mu \nu}F^{\nu}$ is the dual of $F_{\nu}$. The classical field equations are $\nabla^{\mu}F_{\mu}=0$ and $\nabla^{\mu} \,{^{*}F}_{\mu}=0$, where  the scalar field  $\phi$ plays the role of the potential of the field $F_{\mu}$,  $F_{\mu}= \nabla_{\mu} \phi$.  The classical equations are invariant under both Weyl and duality transformation $F_{\mu} \to {^{*}F}_{\mu}$. In this section we consider an arbitrary spacetime metric (not necessarily homogenous), which can always  be written as $ds^2= e^{2\rho} dx^+dx^-$, in terms of the null coordinates $x^{\pm}\equiv t {\pm} x$. Because the spacetime is not necessarily homogenous, we cannot use adiabatic regularization.  We will use instead the Hadamard  point-splitting method \cite{Waldbook, Adler77, Hadamard}, which gives us the chance to show the existence of the duality anomaly  for a different renormalizaton prescription. In this theory there is once again a preferred vacuum state, the conformal vacuum. This state is dual invariant, and so are the vacuum-correlation functions: 
$$\langle F_{{\pm}} (x) F_{{\pm}}(x')\rangle = \frac{-1}{4\pi (x^{\pm} - x'^{\pm})^{2}}= \langle {^{*}F}_{{\pm}} (x) {^{*}F}_{{\pm}}(x')\rangle$$  
$$\langle F_{+} (x) F_{-}(x')\rangle = 0 = \langle {^{*}F}_{+} (x) {^{*}F}_{-}(x')\rangle\, ,$$ 
for $x\neq x'$.  However, for $x=x'$ the subtractions required for renormalization are no longer dual invariant. These subtractions are obtained from the singular part of the  Hadamard two-point function, $1/4\pi  \, [V(x, x') \ln \sigma(x, x')]$, where $2\sigma(x, x')$ is the square of the geodesic distance between $x$ and $x'$ and $V$ is a geometric biscalar \cite{Hadamard, Decanini-Folacci}. We obtain (see Appendix B for details)
%
\bea \label{FF2d}\langle F_{\mu}(x)F_{\nu}(x) \rangle &=& \tilde \theta_{\mu\nu} + \frac{1}{4}\tilde \gamma\, g_{\mu\nu}  \\
\langle {^{*}F}_{\mu}(x){^{*}F}_{\nu}(x) \rangle &=& \tilde \theta_{\mu\nu} - \frac{1}{4}\tilde \gamma\, g_{\mu\nu} \label{FF2d2}\ , \eea
 where $\tilde \theta_{\mu\nu} $ is a traceless tensor with components \cite{Davies-Fulling77} 
 \be \label{theta} \tilde \theta_{{\pm}{\pm}}= -1/12\pi [(\partial_{\pm} \rho)^2-\partial_{\pm}^2 \rho] \, , \ \ \ \tilde \theta_{+-}=0\, , \ee
and  $\tilde \gamma= 1/(12\pi)\, R$. Therefore 
\be \nonumber \tilde \Delta_{\mu\nu}\equiv  \langle F_{\mu}(x)F_{\nu}(x) \rangle - \langle  {^{*}F}_{\mu} (x){^{*}F}_{\nu}(x) \rangle =1/2\, \tilde\gamma \, g_{\mu\nu}\, .\ee 
From (\ref{FF2d}) one can also obtain the vacuum expectation value of the energy-momentum tensor following  the procedure summarized for the electromagnetic case. Taking into account that $\nabla^{\mu}\tilde{\theta}_{\mu\nu}=\frac{1}{48\pi} g_{\sigma\nu}\nabla^{\sigma}R$, one obtains $\langle T_{\mu\nu}\rangle = \tilde \theta_{\mu \nu} - \frac{R}{48\pi}g_{\mu\nu}$, in agreement with \cite{Davies-Fulling77}.

It is well known that for $x\neq x'$ the correlation function $\langle \partial_+\phi (x) \partial_-\phi (x') \rangle$ vanishes, as mentioned before, which is commonly referred to as the decoupling of left- and right-moving modes. A consequence of the duality anomaly is that this is no longer true for $x=x'$. Rather, Eqs.\ (\ref{FF2d}) and (\ref{FF2d2}) provide $\langle \partial_+\phi \partial_-\phi\rangle =\frac{1}{4}\tilde \gamma e^{2\rho}= \frac{1}{ 12\pi} \partial_+\partial_- \rho$.

\section{Conclusions and final comments}
QFT is intrinsically  more involved than a quantum-mechanical system having a finite number of degrees of freedom. This difference arose 
 in the early stages of quantum electrodynamics due to the emergence of divergent expressions in physical quantities.  It was nicely solved with the renormalization program, which has provided many important and surprising results. In particular, when applied in the presence of a classical gravitational background, renormalization  has been shown to break some of the important symmetries of the theory under consideration. The chiral current anomaly for free massless fermions and the Weyl anomaly are examples with important physical consequences. In this paper we have proven that the duality symmetry  cannot  hold in QFT in arbitrarily curved spacetimes.  We have shown this explicitly with some of the most common renormalization methods. However, it could still be  possible to build  a renormalization scheme for which the symmetry is preserved. Even in the case such a method exists, which we believe is unlikely, it would be highly unnatural or fine-tuned.
 
The breakdown of these symmetries mentioned above have a  common origin. The generally covariant singularity structure of the two-point  function only knows about the local properties of the geometry, i.e.\ the metric, curvature tensors and their derivatives. Those singularities do not need to share the symmetries of the theory.
The renormalization process subtracts those local and covariant singularities, and therefore may break the symmetries of the vacuum. This is precisely the case for the duality anomaly discussed here, as can be seen from the renormalization subtractions written in the appendixes.

Phenomenologically, although the duality is an exact physical symmetry of the classical  theory only in the absence of charges, it still plays an important role in certain situations in which charge density is negligible. This happens, for instance, during cosmic inflation. At the conceptual level, electric-magnetic duality has been the focus of several theoretical developments, and an important ingredient in different scenarios, like in the Montonen-Olive dualities \cite{montonen-olive} in non-Abelian gauge and supersymmetric theories. Therefore, the duality anomaly presented in this paper may have interesting physical and theoretical consequences which merit further exploration. \\

\section*{Acknowledgments}   

This work was  supported  by the  Grants. No.\ FIS2011-29813-C02-02, No.\  CPANPHY-1205388, and No.\ MPNS COST Action No. MP1210,  and NSF Grant No.\ PHY-1403943. A.L is supported by the Severo Ochoa program of the Spanish Ministry of Economy and Competitiveness.  I.A. thanks A. Ashtekar, A. Laddha, W. McElgin, and N. Morris  for  discussions. 
J. N.-S.\ thanks L. Parker for the useful comments. 


\appendix
\section{\\ Adiabatic renormalization of $\langle  {F}_{\mu\alpha}{F}_{\nu}^{\ \alpha}\rangle$} \label{AppA}

In this appendix we provide some details of the adiabatic renormalization of the vacuum expectation value $\langle  {F}_{\mu\alpha}(x){F}_{\nu}^{\ \alpha}(x)\rangle$. The formal (unrenormalized) expression can be obtained by using Eq.\ (\ref{A}). One obtains

\bea \label{vac}  \langle  {F}_{\mu\alpha}(x){F}_{\nu}^{\ \alpha}(x)\rangle   
 =  -g_{\mu\nu} \frac{1}{ \pi^2 a^4(\eta)}\int_0^{\infty} dk\, k^2 \, \frac{k}{2}   \, .\eea
Note that the same formal integral is obtained for $\langle  {^{*}F}_{\mu\alpha}(x){^{*}F}_{\nu}^{\ \alpha}(x)\rangle$. This integral diverges as the fourth power of the comoving momentum $k$. In adiabatic regularization, and also in the DeWitt-Schwinger  method, to find the renormalization subtraction terms that make the above integral finite, one has to first temporarily introduce a mass in the theory, then take the limit $m\to 0$ at the end of the calculation.  On top of that one has to introduce also the familiar gauge breaking term $-1/2(\nabla _{\mu} A^{\mu})^2$, ghost term $\nabla_{\mu}c\nabla^{\mu}c^*$, and the corresponding (temporary) mass terms
 $1/2m^2 A_{\mu}A^{\mu} - m^2c^*c$ (see, for instance, \cite{Christensen, birrell-davies}).
The ghost field 
 is required to maintaing gauge invariance when taking $m\to 0$.
  The introduction of a temporary mass is a fundamental requirement in the adiabatic and DeWitt-Schwinger methods \cite{parker-fulling, Christensen}. The role of the auxiliary mass is to avoid the emergence of artificial infrared divergences when timing the UV ones. Note that the DeWitt-Schwinger expansion of the Feynman propagator is an asymptotic  expansion in inverse powers of $m^2$ \cite{DeWitt75}.

Therefore, the adiabatic expansion of the vector potential,  $A^{\rm (Ad)}_{\mu}$, contains two transverse and one longitudinal polarization. The  expansion of both polarizations up to fourth adiabatic order provides the subtraction terms needed to renormalize the expectation values we are looking for. Note that one must include terms of up to fourth adiabatic order in the subtractions because this is the order at which divergences appear for  generic spacetime metrics, not necessarily conformally flat, and for general values of the mass. In the $m\to 0$ limit there are no divergences at fourth order, but one still must apply the general prescription.

The transverse polarizations take the same form as in Eq.\ (\ref{A}), with the only difference that  now the mode functions $\varphi^{\rm (Ad)}_k(\eta)$  satisfy the equation 
\be\partial^2_{\eta}\varphi^{\rm (Ad)}_k+\omega(k,\eta)^2 \varphi^{\rm (Ad)}_k=0\, , \ee  
with $\omega(k,\eta)=\sqrt{k^2+m^2a(\eta)^2}$. This expression  is identical to the equation satisfied by the modes of a scalar field conformally coupled to the FLRW metric. The adiabatic expansion of $\varphi^{\rm (Ad)}_k$, up to fourth order, is
\be \varphi^{\rm (Ad)}_k(\eta)=\frac{1}{ \sqrt{2 W_{\rm trs}(k,\eta)}}e^{-i \int^{\eta} d\eta' \, W_{\rm trs}(k,\eta') }\, ,\ee
with $W_{\rm trs}(k,\eta)=W_0+W_1+W_2+W_3+W_4$, where
\bea 
W_0&=&w(k,\eta)\nonumber\\
W_2&=&\frac{2 w'^2-2w w''}{8 w^3}\nonumber\\
W_4&=&\frac{1}{128w^7}(-297 w'^4-396 w w'^2w''-52w^2w''^2\nonumber\\&-&80w^2w'w'''+8w^3w'''')\nonumber\\
W_1&=&W_3=0\, .\eea
The prime in the previous equations indicates derivative with respect to conformal time.

 The longitudinal polarization of $A^{\rm (Ad)}_{\mu}$ can be chosen as $\epsilon^{(3)}_{\mu} \chi_k^{\rm (Ad)}(\eta)$, where $\epsilon^{(3)}_{\mu}$ has components 
\be \epsilon^{(3)}_{\mu}(\vec{k})=\left(\begin{array}{c}
f(k,\eta)\\
k_{1}\\
k_{2}\\
k_3
\end{array}\right) ; \,  \, \, f(k,\eta)= -i \frac{k^2}{\omega^2} \frac{\partial_{\eta}\chi^{\rm (Ad)}_k}{\chi^{\rm (Ad)}_k}\, ,\ee
%
and the rescaled mode $\psi_k(\eta)=\frac{k \, m}{\omega} \chi_k^{\rm (Ad)}(\eta)$ satisfies the equation
\be \psi_k''+2\frac{a'}{a} \psi'_k+\left[\frac{\omega''}{\omega}+2\frac{a'}{a} \frac{\omega'}{\omega}-2 \left(\frac{\omega'}{\omega}\right)^2+\omega^2\right] \psi_k=0\, .\ee
The adiabatic expansion of $\psi_k(\eta)$ is given by 
\be \psi_k(\eta)=\frac{1}{a(\eta) \sqrt{2 W_{\rm long}(k,\eta)}}e^{-i \int^{\eta} d\eta' \, W_{\rm long}(k,\eta') }\, ,\ee
with $W_{\rm long}(k,\eta)=\tilde W_0+\tilde  W_1+\tilde  W_2+\tilde  W_3+\tilde W_4$, where
\bea 
\tilde W_0&=&w(k,\eta)\nonumber\\
\tilde W_2&=&\frac{8wa'w'-5aw'^2-4w^2a''+2aww''}{8 w^3}\nonumber\\
\tilde W_4&=&\frac{1}{128a^3 w^7}(-64w^3a'^3w'-288aw^2a'^2w'^2\nonumber\\&-&400 a^2wa'w'^3+455a^3w'^4+32w^4a'^2a''\nonumber\\&+&240aw^3a'w'a''+312 a^2w^2w'^2a''-32 a w^4a''^2\nonumber\\&+&64aa'^2w^3w''+288a^2w^2a'w'w''-540a^3ww'^2w''\nonumber\\&-&80a^2w^3 a''w''+60 a^3w^2w''^2-32aw^4a'a'''\nonumber\\&-&112a^2w^3w'a'''-32a^2w^3a'w'''+96a^3w^2w'w'''\nonumber\\&+&16a^2w^4a''''-8a^3w^3w'''')\nonumber\\
\tilde  W_1&=&\tilde W_3=0\, .\eea

By substituting the adiabatic modes in the expression for  $\langle  {F}_{\mu\alpha}{F}_{\nu}^{\ \alpha}\rangle$ and keeping terms up to fourth adiabatic order, one obtains the renormalization subtraction terms. As an example, the renormalized expression of the time-time component has the form (recall we work here in conformal time)

\bea \label{adF}& &\langle  {F}_{0\alpha}{F}_{0}^{\ \alpha}\rangle=\\&-&\frac{a'^4-14 a a'^2 a''+4 a^2a''^2+12a^2a'a'''-5a^3 a''''}{480\pi^2 a^6}+\nonumber \\ &&\frac{-58a'^4+122aa'^2a''-36a^2a'a'''+a^2(-17 a''^2+4 a a'''')}{480\pi^2a^6}\nonumber \, .\eea
In the above expression the transverse adiabatic modes have exactly canceled the quartic divergence of the vacuum contribution (\ref{vac}), and provide, additionally, the finite term showed in the second line of the previous equation. The third line of that equation is the contribution of the longitudinal adiabatic polarization. Expression (\ref{adF}) agrees with the time-time component of Eq.\ (\ref{FF}) for $t_{\mu\nu}=0$. The rest of the components are computed in the same way.  In sharp contrast, the longitudinal adiabatic modes make no contribution to $\langle  {^{*}F}_{0\alpha}(x){^{*}F}_{0}^{\ \alpha}(x)\rangle$. This explains the difference between (\ref{FF}) and (\ref{*F*F}).


\section{\\Hadamard renormalization of $\langle  {F}_{\mu}{F}_{\nu}\rangle$} \label{AppB}

In Hadamard renormalization (for details of the specific Hadamard prescription used here the reader is referred to \cite{Hadamard, Decanini-Folacci})
the physically relevant, finite expectation values are obtained as
\be  \label{Hadren}  \langle  {F}_{\mu}(x){F}_{\nu}(x)\rangle=\lim_{x'\to x}\, \nabla_{\mu}\nabla_{\nu'}[\langle  \phi(x)\phi(x') \rangle-H_{\rm sing}(x,x')] \, . \ee
In this equation $\nabla_{\nu'}$ indicates the covariant derivative with respect to $x'$.  $H(x,x')$ is a bidistribution with  Hadamard's-type singularity structure, which in $1+1$ dimensions takes the form \cite{Decanini-Folacci}
\be H_{\rm sing}(x,x')=\frac{1}{4\pi}[V(x,x') \ln \sigma(x,x')] \, , \ee
where $\sigma(x,x')$ is half of the square of the geodesic distance between the points $x$ and $x'$, and $V(x,x')$ is a biscalar which admits an expansion of the form
\be V(x,x')= \sum_{n=0}^{\infty} V_n(x,x')\, \sigma^n\, .\ee
The field equations provide recursion relations which uniquely determine the coefficients $V_n(x,x')$ from $V_0(x,x')=-\Delta(x,x')^{1/2}$, where $\Delta(x,x')$ is the Van Vleck-Morette determinant (see e.g. \cite{birrell-davies}).

The vacuum two-point function of the field $\phi(x)$ takes the same form as in Minkowski spacetime, due to the conformal symmetry of the  field equations. In terms of the null coordinates that were introduced in Sec. (\ref{2D}), it reads  
\be \label{2pf} \langle  \phi(x)\phi(x') \rangle=-\frac{1}{4\pi}\ln|(x^+-x'^+)(x^--x'^-)|\,. \ee
This correlation function can be written in Hadamard form  
\be\nonumber \langle  \phi(x)\phi(x') \rangle= \frac{1}{4\pi}[V(x,x')\ln \sigma(x,x')+W_{\rm conf}(x,x')]\,, \ee
where $W_{\rm conf}(x,x')= \omega(x) + \omega_{\mu}(x)\sigma^{;\mu} + 1/2! \, \omega_{\mu\nu}(x) \sigma^{;\mu}\sigma^{;\nu} + ...$  is the biscalar that encodes the state dependence of the  two-point function. Substituting in Eq.\ (\ref{Hadren}), we have 
\bea  \label{H}  \langle  {F}_{\mu}(x){F}_{\nu}(x)\rangle&=&\frac{1}{4\pi}\lim_{x'\to x}\,  \nabla_{\mu}\nabla_{\nu '}[W_{\rm conf}(x,x')] \nonumber \\ 
&=&\frac{1}{4\pi}(-\omega_{\mu\nu}+\frac{1}{2} \, \omega_{;\mu \nu})\, . \eea
In our case, $\omega_{\pm\pm}= 2/3 \, \partial^2_{\pm} \rho - 5/3 \, (\partial_{\pm} \rho)^2$, $\omega_{+-}=2/3 \, \partial_+\partial_- \rho$ and $\omega = 2\rho$. Hence, one obtains 
\be  \langle  {F}_{\mu}(x){F}_{\nu}(x)\rangle=\lim_{x'\to x}\,  \nabla_{\mu}\nabla_{\nu}[W_{\rm conf}(x,x')]=\tilde \theta_{\mu\nu}+\frac{R}{48 \pi} \, g_{\mu\nu}\ee
where $\tilde \theta_{\mu\nu}$ was defined in (\ref{theta}). This is the result shown in Eq.\ (\ref{FF2d}). From this it is very easy to get (\ref{FF2d2}).


\begin{thebibliography}{99}
\bibitem{note1} We will refer to  these metric rescalings as Weyl transformations, to distinguish them from changes of coordinates whose effect is to change the metric tensor by a multiplicative positive factor. These conformal diffeomorphisms are often called conformal transformations in the literature \cite{CFT}.
\bibitem{note2} Weyl-invariant theories in  curved spacetime are also invariant under conformal coordinate transformations in the flat-space limit. This conformal invariance has been shown to break down for free fields in two-dimensional flat spacetime due to renormalization effects \cite{Fulling-Davies76, Davies-Fulling77}. In some sense this conformal anomaly is  more analogous to the duality symmetry breaking studied here.
\end{thebibliography}
\end{document}